\documentclass[doublecol]{epl2mod} 

\definecolor{asparagus}{rgb}{0.53, 0.66, 0.42}
\definecolor{cadmiumgreen}{rgb}{0.0, 0.42, 0.24}
\definecolor{question}{rgb}{0.7, 0.6, 0.1}

\newcommand{\ff}[1]{{\boldsymbol #1}}

\newcommand{\ca}[1]{{\cal #1}}

\newcommand{\tr}{\mbox{tr} \,}
\newcommand{\scl}{S_{\rm cl.}}

\newcommand{\bi}{\begin{itemize}}
\newcommand{\ei}{\end{itemize}}
\newcommand{\be}{\begin{equation}}
\newcommand{\ee}{\end{equation}}

\newcommand{\parag}[1]{{\em #1}--}

\begin{document} 

\title{
Inertia effects in the real-time dynamics of a quantum spin coupled to a Fermi sea
}

\author{Mohammad Sayad \and Roman Rausch \and Michael Potthoff}
\institute{I. Institut f\"ur Theoretische Physik, Universit\"at Hamburg, Jungiusstra\ss{}e 9, 20355 Hamburg, Germany}




\abstract{
Spin dynamics in the Kondo impurity model, initiated by suddenly switching the direction of a local magnetic field, is studied 
by means of the time-dependent density-matrix renormalization group. 
Quantum effects are identified by systematic computations for different spin quantum numbers $S$ and by comparing with tight-binding spin-dynamics theory for the classical-spin Kondo model.
We demonstrate that, besides the conventional precessional motion and relaxation, the quantum-spin dynamics shows nutation, similar to a spinning top.
Opposed to semiclassical theory, however, the nutation is efficiently damped on an extremely short time scale. 
The effect is explained in the large-$S$ limit as quantum dephasing of the eigenmodes in an emergent two-spin model that is weakly entangled with the bulk of the system. 
We argue that, apart from the Kondo effect, the damping of nutational motion is essentially the only characteristics of the quantum nature of the spin. 
Qualitative agreement between quantum and semiclassical spin dynamics is found down to $S=1/2$.
}

\maketitle 

\parag{Introduction.} 
The paradigmatic system to study the real-time dynamics of a spin-$1/2$ coupled to a Fermi sea is the Kondo model \cite{Kon64}. 
It is mainly considered as a generic model for the famous Kondo effect \cite{Hew93}, namely screening of the impurity spin by a mesoscopically large number of electrons in a thermal state with temperature below the Kondo temperature $T_{\rm K} \sim \exp(-1 / J \rho)$, where $J$ is the strength of the exchange coupling and $\rho$ is the density of states.
The Kondo effect is a true {\em quantum} effect which originates from the two-fold spin degeneracy
and is protected by time-reversal symmetry.
Longitudinal spin dynamics, such as the time-dependent Kondo screening, has been studied recently \cite{MHK13,NGA+15}
by starting from an initial state with a fully polarized spin, which can be prepared with the help of local magnetic field. 
The longitudinal dynamics is initiated by suddenly switching off the field.

Transversal spin dynamics, on the other hand, appears as a more {\em classical} phenomenon: 
It can be induced, for example, by suddenly tilting a strong field $B \gg T_{\rm K}$ from, say, $\hat{x}$ to $\hat{z}$ direction.
In first place this induces a precession of the spin around the new field direction with Larmor frequency $\omega_{\rm L} \propto B$.
For $J=0$, the equation of motion for the expectation value of the spin, $(d/dt) \langle \ff S \rangle_{t} = \langle \ff S \rangle_{t}\times \ff B$ with $\ff B = B \hat{z}$ has the same form as the Landau-Lifschitz equation for a classical spin \cite{LL35}.
When coupling the spin to the Fermi sea with a finite $J$, energy can be transferred to the electronic system and dissipated into the bulk.
Hence, the spin must relax and align to the new field direction as is nicely seen in numerical studies of the Kondo model out of equilibrium \cite{AS06}.
For $B \gg T_{\rm K}$ the spin precession and relaxation is qualitatively well described by semiclassical tight-binding spin dynamics (TB-SD) (cf.\ e.g.\ Ref.\ \cite{SP15}) where the spin is assumed to be a classical dynamical observable. 
In many cases, even the simple Landau-Lifschitz-Gilbert (LLG) equation \cite{Gil55,Gil04} including a non-conserving damping term, proportional to the first time derivative of the spin, 
seems to capture the essential (classical) physics.

A major purpose of the present study is to check if there are quantum effects which are overlooked by the semiclassical approach to transversal spin dynamics (i.e., apart from the Kondo effect). 
To this end we compare numerical results from exact quantum-classical hybrid theory \cite{Elz12,Sal12}, i.e., the TB-SD \cite{SP15,SRP16}, with those of exact quantum theory, computed with time-dependent density-matrix renormalization group (t-DMRG) \cite{Sch11,HCO+11}, for different spin quantum numbers $S$.
It turns out that even for $S=1/2$ there is a surprisingly good qualitative agreement of quantum with semiclassical dynamics.
However, we also identify a physical phenomenon, namely nutational motion, where remarkable differences are found:

\parag{Classical and quantum nutation.}
Besides precession and damping, inertia effects are well known in classical spin dynamics \cite{But06,WC12} and can be described by an additional term to the LLG equation with second-order time derivative of the spin.
The resulting nutation of the spin motion has been introduced and studied phenomenologically \cite{CRW11,OLW12} or with realistic parameters taken from first-principles calculations \cite{BH12} but can also be derived on a microscopic level \cite{FSI11,BNF12,KT15} within the general framework of semiclassical spin dynamics \cite{TKS08,FI11,EFC+14}.

In case of a quantum spin, inertia effects have not yet been studied. 
As compared to spin precession and damping, nutation is a higher-order effect \cite{BNF12}, so that it is not {\em a priori} clear whether or not spin nutation is suppressed by quantum fluctuations.
Here, by applying the t-DMRG to the spin-$S$ Kondo impurity model in a magnetic field, we are able to show for the first time that nutation also shows up in the full quantum spin dynamics. 
Remarkably, however, {\em quantum nutation} turns out to be strongly damped and shows up on a much shorter time scale as compared to the relaxation time.
On a fundamental level, this pinpoints an unconventional new quantum effect in transversal spin dynamics but is also relevant for experimental studies suggesting, e.g., inertia-driven spin switching \cite{KIP+09,KKR10} opposed to standard precessional switching \cite{GvdBH+02,TSK+04}.

\parag{Model.}
Using standard notations, the Hamiltonian of the Kondo impurity model reads:
\be
H = - T \sum_{i<j}^{n.n.} \sum_{\sigma = \uparrow,\downarrow} (c_{i\sigma}^{\dagger} c_{j\sigma} + \mbox{H.c.})
+ J \ff s_{i_0} \ff S 
- \ff B \ff S \; .
\label{eq:ham}
\end{equation}
Here, $c_{i\sigma}$ is the annihilator of an electron with spin projection $\sigma=\uparrow,\downarrow$ at site $i=1,...,L$ of an open one-dimensional chain of length $L$. 
The hopping $T=1$ between nearest-neighboring (n.n.) sites defines the energy and the time scale ($\hbar \equiv 1$). 
We assume a half-filled band with $N=L$ conduction electrons.
The impurity spin $\ff S$ is coupled antiferromagnetically with exchange coupling constant $J$ to the local spin $\ff s_{i_{0}}$ of the itinerant conduction-electron system at the first site of the chain, $i_{0}=1$.
With the vector of Pauli matrices $\ff \tau$, we have $\ff s_{i} = \sum_{\sigma\sigma'} c^{\dagger}_{i\sigma} \ff \tau_{\sigma\sigma'} c_{i\sigma'} / 2$.

$\ff S$ is a quantum spin characterized by quantum number $S = \frac12,1,\frac32, ...$, and for $S>1/2$, Eq.\ (\ref{eq:ham}) is the underscreened Kondo model.
Alternatively, $\ff S$ is considered as a classical spin with fixed length $|\ff S| = \scl$ where $\scl = \sqrt{S(S+1)}$ for a meaningful comparison with results for a quantum spin.

\parag{Real-time dynamics.}
To initiate spin dynamics we consider a local magnetic field $\ff B$ which, at time $t=0$, is suddenly switched from $\ff B = B_{\rm ini} \hat{x}$, forcing the spin to point in $\hat{x}$ direction, to $\ff B = B_{\rm fin} \hat{z}$.
This addresses, e.g., spin-resolved scanning-tunneling microscope experiments \cite{Wie09,NF93,Mor10,LEL+10,YCB+14}.
We choose $B_{\rm ini} = \infty$ to initially fully polarize the impurity spin.
Note that the conduction-electron spin $\ff s_{i_{0}}$ in the initial state is also polarized, but typically much weaker, depending on the internal Weiss field $\ff B_{\rm eff} \equiv J\ff S$ produced by the exchange interaction and the impurity spin.
The dynamics is (predominantly) transversal if $B_{\rm fin} \gg T_{\rm K}$ which ensures that the Kondo singlet remains broken and that there are no (significant) longitudinal spin fluctuations.

For $t\to \infty$ we expect complete relaxation. 
This is achieved if the classical spin $\ff S(t)$ or, in the quantum case, $\ff S(t) \equiv \langle \ff S \rangle_{t} = \langle \Psi(t) | \ff S | \Psi(t) \rangle$ fully aligns with the $\hat{z}$ axis. 
Likewise the expectation value $\ff s_{i_{0}}(t) \equiv \langle \ff s_{i_{0}} \rangle_{t}$ of the local conduction-electron spin at $i_{0}$ is expected to orient itself antiparallel to $\ff S(t)$ for $t\to \infty$.

\parag{Time-dependent DMRG.}
To study the (quantum) time-evolution of $\ff S(t)$ and $\ff s_{i_{0}}(t)$ after the sudden switch of the field, we employ the time-dependent density-matrix renormalization-group technique (t-DMRG) in the framework of matrix-product states and operators \cite{Sch11}.
The implementation of a quantum spin with arbitrary $S$ is straightforward. 
For an impurity model with the spin attached to the first site of the chain, the numerical effort is essentially independent of $S$ as only the dimension of the local Hilbert space at $i_{0}$ scales with $2S+1$.
Due to the global $U(1) \times U(1)$ symmetry of $H$, the total particle number and the $z$ component of the total spin are conserved. 
For a sudden field switch from $\hat{x}$ to $\hat{z}$ direction, however, only particle-number conservation can be exploited in the t-DMRG calculation. 
As compared to a purely longitudinal dynamics, this implies an increased computational effort.
The time evolution of matrix-product states is computed using the two-site version of the algorithm as suggested in Ref.\ \cite{HCO+11,HLO+14} which is based on the time-dependent variational principle. 
The maximum bond dimension reached during the propagation is about 2000.

\begin{figure}[t]
\centering
\includegraphics[width=0.95\columnwidth]{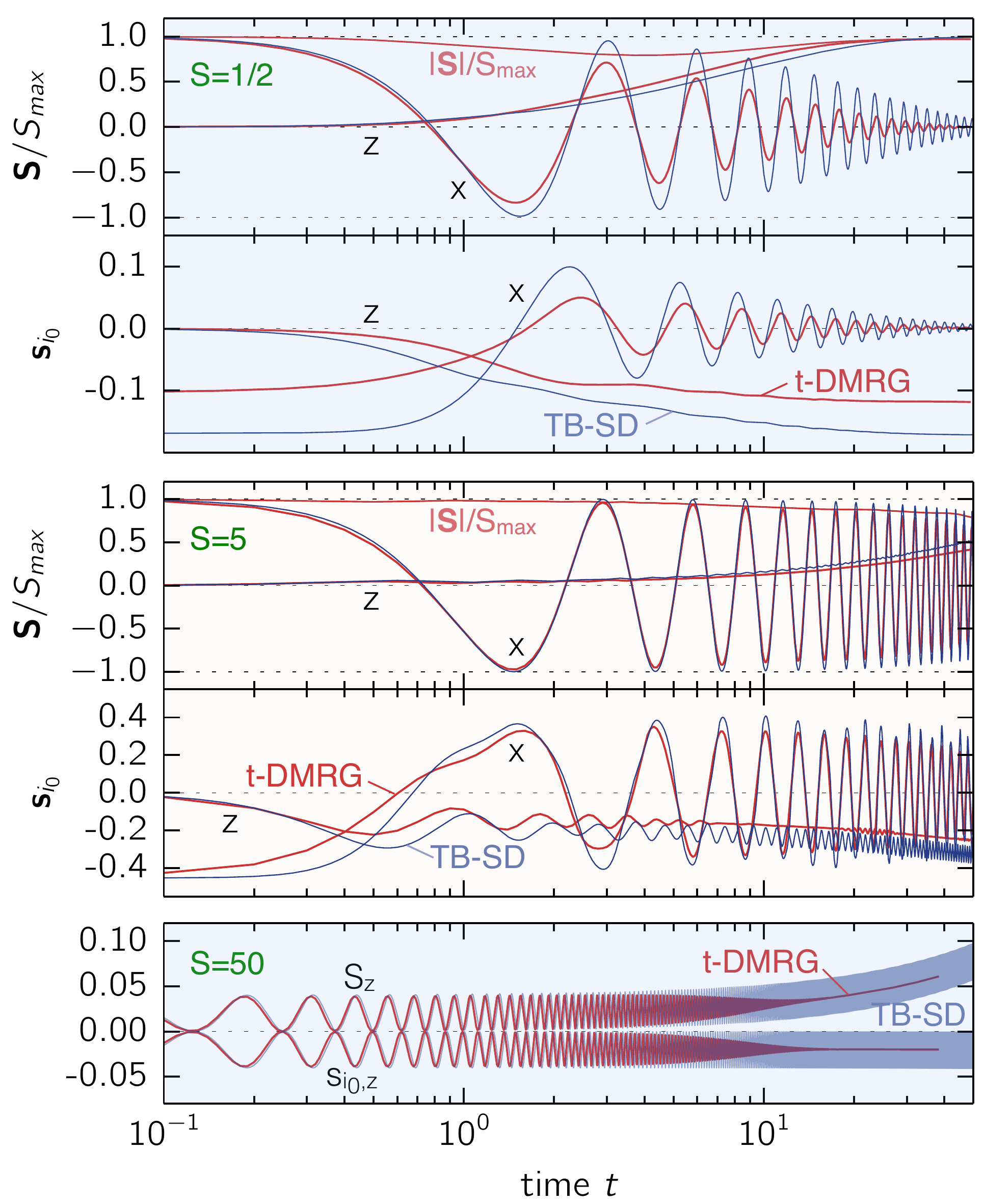}
\caption{
{\em Top panel, upper part:}
Dynamics of $\ff S(t) / S_{\rm max}$ for the Kondo impurity model, Eq.\ (\ref{eq:ham}), for $J=1$ and $\ff B = B_{\rm fin} \hat{z}$ with $B_{\rm fin}=2$. Only $x$ and $z$ components are shown.
At $t=0$, the system is prepared with $\ff S(0) / |\ff S(0)| = \hat{x}$.
Time units are fixed by the inverse hopping $1/T \equiv 1$. 
Red lines: t-DMRG calculations for a quantum spin, $\ff S(t) \equiv \langle \Psi(t) | \ff S | \Psi(t) \rangle$, and $S=1/2$ ($S_{\rm max}=S$).
Blue lines: semiclassical dynamics (TB-SD) with a classical spin $\ff S(t)$ of length $\scl = \sqrt{S(S+1)} = \sqrt{3}/2$ ($S_{\rm max}=\scl$).
{\em Top panel, lower part:}
Local conduction-electron moment $\ff s_{i_{0}}(t) \equiv \langle \ff s_{i_{0}} \rangle_{t}$.
{\em Middle}:
The same for $S=5$.
{\em Bottom}:
$z$ components of $\ff S(t)$ and $\ff s_{i_{0}}(t)$ for $S=50$.
} 
\label{fig:dyn}
\end{figure}

\parag{Quantum-spin dynamics.}
We start the discussion with the t-DMRG results, see the red lines in Fig.\ \ref{fig:dyn}. 
The calculations have been performed for a chain with $L=80$ sites. 
For a quantum spin $S=1/2$ (Fig.\ \ref{fig:dyn}, top panel), and for $J=1$ and $B_{\rm fin} = 2$, the dynamics is sufficiently fast, i.e., the main physical effects take place on a time scale shorter than the time where finite-size artifacts show up.
In the bulk of the non-interacting conduction-electron system, wave packets typically propagate with group velocity
$v_{\rm F} = d \varepsilon(k) / dk = \pm 2 T$ at the Fermi wave vectors $k=k_{\rm F} = \pm \pi / 2$ for half filling.
This roughly determines the maximum speed of the excitations and defines a ``light cone'' \cite{LR72,BHV06}.
Hence, a local perturbation at $i_{0}=1$ starts to show artificial interference with its reflection from the opposite boundary at $i=L$ after a time of about $t_{\rm inter} = 2L / v_{\rm g} = L/T$, i.e., after about 80 inverse hoppings -- which is well beyond the time scale covered by Fig.\ \ref{fig:dyn}.

The most obvious effect in the time dependence of $\ff S(t)$ (see upper part of the top panel) is the precessional motion around the $\hat{z}$ axis: 
$S_{x}(t)$ (and likewise of $S_{y}(t)$ which is not shown in the figure) oscillate with Larmor frequency $\omega_{\rm L} \approx B_{\rm fin}$.
Note that $|\ff S(t)| = | \langle \Psi(t) | \ff S | \Psi(t) \rangle |$ is nearly constant, i.e., there are no substantial longitudinal fluctuations or Kondo screening.

In addition to the spin precession, there is damping:
The spin relaxes to its new equilibrium direction $\propto \hat{z}$ on the relaxation time scale $\tau_{\rm rel} \approx 50$.
Despite the fact that the total energy and the $z$ component of the total spin are conserved (as is also checked numerically), this is the expected result: 
At $t=0$ the system is locally in an excited state; for large $t$, spin relaxation is achieved by dissipation of energy into the bulk of the  chain.
The dynamics does not stop until the excitation energy $\sim S B_{\rm fin}$ is fully dissipated into the bulk, and the system is -- locally, close to $i_{0}$ -- in its ground state.

\parag{Conduction-electron dynamics.}
In the ground state of the system at time $t=0$, the local conduction-electron spin at $i_{0}$ is partially polarized  in $-\hat{x}$ direction, i.e., antiparallel to $\ff S(t=0)$ due to the internal magnetic field $J \ff S(0)$ (see top panel of Fig.\ \ref{fig:dyn}, lower part).
For $t>0$ we find that $\ff s_{i_{0}}(t)$ follows the dynamics of the impurity spin $\ff S(t)$ {\em almost} adiabatically, i.e., at a given instant of time $t$ it is slightly behind the (instantaneous) ground-state expectation value $\langle \ff s_{i_{0}} \rangle_{\rm g.s.} \uparrow\downarrow \ff S(t)$ for the conduction-electron system with a ``given'' Weiss field $J \ff S(t)$.
This slight retardation effect is clearly visible in Fig.\ \ref{fig:dyn} (compare the location of the first minimum of $S_{x}(t)$ with the first maximum of $s_{i_{0}x}(t)$, for instance).
In the semiclassical picture retardation has been identified to drive the relaxation of $\ff S(t)$ \cite{SP15}.

\parag{Quantum nutation.}
In addition to the expected precessional motion and relaxation of $\ff s_{i_{0}}(t)$, there is a weak additional superimposed oscillation visible in $s_{i_{0}z}(t)$. 
For $S=1/2$ the frequency is close to the precession frequency.
However, the results for higher spin quantum numbers (see lower part of the middle panel, $S=5$) show that these oscillations have a characteristic frequency $\omega_{\rm N}$ and hence a physical cause which may require but is independent of the precessional motion.

The $z$ component of the impurity spin actually shows oscillations with the same frequency and almost the same amplitude (which can hardly be seen in the first two panels of Fig.\ \ref{fig:dyn} due to the rescaling of $\ff S(t)$ by $S_{\rm max}$) but becomes obvious in the bottom panel (no rescaling, $S=50$).
By comparing with the semiclassical spin dynamics, we will argue that this is in fact nutation of the quantum spin.

\parag{Tight-binding spin dynamics.}
Most (but not all) features of the transversal quantum dynamics are qualitatively captured by the numerically much cheaper ``tight-binding spin dynamics'' (TB-SD) \cite{SP15,SRP16}, i.e., quantum-classical hybrid or Ehrenfest dynamics.
TB-SD originates from the Hamiltonian Eq.\ (\ref{eq:ham}) by treating the impurity spin $\ff S(t)$ as a classical dynamical observable which couples to the (quantum) system of conduction electrons.
Its equation of motion is derived from the canonical equation $\dot{\ff S} = \{ \ff S, \langle H \rangle_{t} \}$ (see Refs.\ \cite{Elz12,SP15} for the Poisson bracket of spin systems), which has the form of a Landau-Lifschitz equation, 
\be
\dot{\ff S}(t) = \ff S(t) \times \ff B  - J \ff S(t) \times \ff s_{i_{0}}(t)  \: .
\label{eq:eoms}
\ee
To also get $\ff s_{i_{0}}(t) = \frac12 \tr_{2\times 2} \ff \rho_{i_{0}i_{0}}(t) \ff \sigma$, it must be complemented, however, by a von Neumann equation, $i \frac{d}{dt} \ff \rho(t) = [ \ff T(t) , \ff \rho(t) ]$, 
for the reduced one-particle density matrix $\ff \rho(t)$ of the electron system whose elements are defined as
$\rho_{ii',\sigma\sigma'}(t) \equiv \langle c_{i'\sigma'}^{\dagger} c_{i\sigma} \rangle_{t}$. 
Here, the elements of the effective hopping matrix are $T_{ii',\sigma\sigma'}(t) = -T \delta_{\langle ii' \rangle} \delta_{\sigma\sigma'} + \delta_{ii_{0}} \delta_{i'i_{0}} \frac{J}{2} (\ff S(t) \ff \sigma)_{\sigma\sigma'}$.
The numerical solution using a high-order Runge-Kutta method is straightforward \cite{Ver10}.

\parag{Results of the semiclassical approach.}
TB-SD results are shown by light blue lines in Fig.\ \ref{fig:dyn}.
To make contact with the t-DMRG data, we again consider $L=80$ sites although much larger systems could be treated numerically (see for instance Ref.\ \cite{SP15}).
Overall, the semiclassical theory produces qualitatively very similar results as compared to the quantum dynamics.
This concerns the precessional motion, the relaxation time scale and also the occurrence of nutation and the nutation frequency and amplitude.

However, we can identify basically three quantum effects which are different or even absent in the TB-SD:

(i) Initially the local conduction-electron spin at $i_{0}$ is less polarized in the quantum case, and this has some quantitative consequences for the subsequent spin dynamics. 
The reason is that with $\scl = \sqrt{S(S+1)}$ the classical Weiss field is stronger: $J S_{\rm cl} = J \sqrt{3}/2 > J/2 = J S$.

(ii) Opposed to the classical-spin case, which exclusively comprises transversal dynamics, we find $|\ff S(t)| \ne \mbox{const}$ in the quantum case, i.e., there are residual longitudinal fluctuations (see top panel, upper part). 
Due to the suppression of the Kondo effect by the magnetic field, these are moderate, such that the deviations from the TD-SD are small. 
One should note, however, that nevertheless (weak) longitudinal fluctuations are {\em essential} for true quantum spin dynamics:
Assuming the complete absence of longitudinal fluctuations, we would have $\langle \ff S \rangle_{t} = S \, \hat{n}(t)$ with some unit vector $\hat{n}(t)$. 
Aligning the momentary quantization axis to $\hat{n}(t)$, the quantum state at time $t$ is a product state with zero impurity-bath entanglement.
For the impurity-spin equation of motion, $d \langle \ff S \rangle_{t} / dt = \langle \ff S \rangle_{t} \times \ff B - J \langle \ff S \times \ff s_{i_{0}} \rangle_{t}$, this implies the factorization $\langle \ff S \times \ff s_{i_{0}} \rangle_{t} =  \ff S(t) \times \ff s_{i_{0}}(t)$, resulting in Eq.\ (\ref{eq:eoms}). 
With the analogous factorization in the equations of motion for the conduction-electron degrees of freedom, this implies classical spin behavior. 
Hence, longitudinal fluctuations produce entanglement and quantum effects.

(iii) The nutational motion is strongly damped in the quantum-spin case.
Oscillations of $S_{z}(t)$ and of $s_{i_{0}z}(t)$ with frequency $\omega_{\rm N}$ decay on a finite time scale $\tau_{\rm N}$ while there is no visible damping of the nutation for a classical spin on the scale displayed in Fig.\ \ref{fig:dyn}.
This is most obvious for $S=50$ (bottom panel), but also for $S=5$ (middle panel, lower part).

\begin{figure}[t]
\centering
\includegraphics[width=0.9\columnwidth]{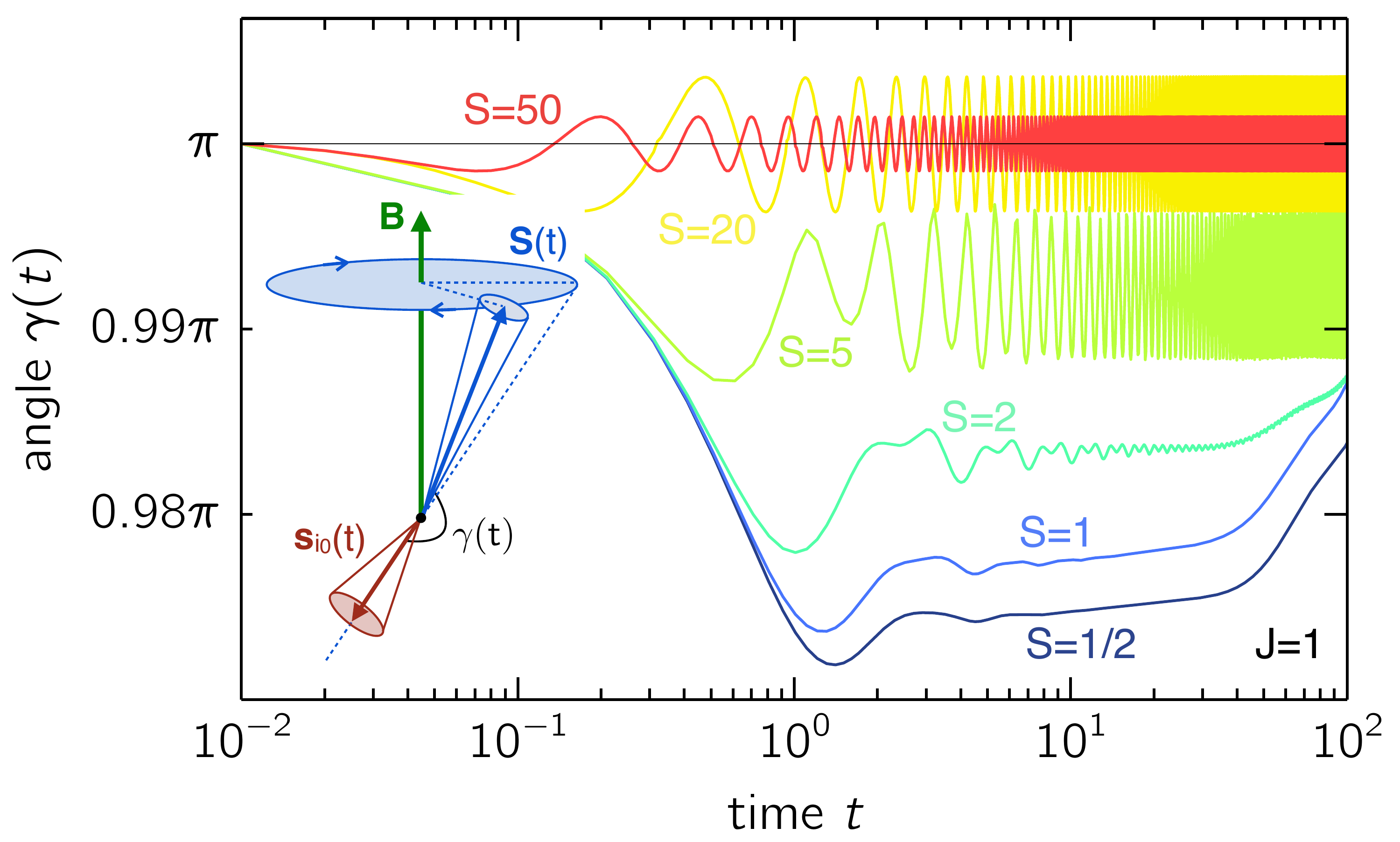}
\caption{
Angle $\gamma(t)$ between $\ff S(t)$ and $\ff s_{i_{0}}(t)$ in the spin dynamics after the sudden switch of the field from $\hat{x}$ to $\hat{z}$ direction. 
TB-SD results for $J=1$, $B_{\rm fin} = 0.1$ and different $\scl = \sqrt{S(S+1)}$ as indicated.
{\em Inset:} schematic illustration of the nutational motion, see text.
} 
\label{fig:angle}
\end{figure}

\parag{$S$ dependence.}
For large spin quantum numbers, one expects that the quantum-spin dynamics becomes equivalent with that of a classical spin of length $\scl = \sqrt{S(S+1)}$ \cite{Lie73,KFG99,GKF00,Gar08,SGP12b}.
Indeed, the agreement constantly improves with increasing $S$, see Fig.\ \ref{fig:dyn}.
The common trends found with increasing $S$ are the following:

(i) There is a stronger and stronger initial polarization of the local conduction-electron spin at $i_{0}$ due to the increasing magnitude of the Weiss field $\ff B_{\rm eff} \equiv J \ff S$ coupling to $\ff s_{i_{0}}$. 
For $S=5$ it is more than 80\% polarized. 

(ii) The relaxation time $\tau_{\rm rel}$ increases with increasing $S$.
For $S=5$ (see Fig.\ \ref{fig:dyn}, middle panel) $S_{z}(t)$ has reached only 50\% of its final saturation value, and for $S=50$ (bottom panel) there is hardly any damping visible on the time scale accessible to the t-DMRG computations. 
Within weak-$J$ perturbation theory and assuming that the spin dynamics is slow as compared to the electronic time scales, we expect $\tau_{\rm rel} \propto S$ in the large-$S$ limit, as is detailed in the Supplemental Material \cite{suppl}.
However, for both the semiclassical and the quantum theory, we find $\tau_{\rm rel} \propto S^{2}$ from the data. 
This is at variance with LLG theory and can be traced back to the breakdown of the Markov approximation (see \cite{suppl}).

(iii) For the nutation frequency we find $\omega_{\rm N} \propto S$ in the large-$S$ limit (see also the discussion below).
The amplitude of the nutation vanishes for $S\to \infty$ in both, the quantum- and the classical-spin case. 
In this way quantum- and classical-spin dynamics become equivalent in the large-$S$ limit despite the absence of damping of the nutational motion in the classical case. 

(iv) We finally note that $|\ff S(t)| / S_{\rm max}$ becomes constant in the quantum case as $S \to \infty$. 

\parag{Microscopic cause of nutation.}
The nutational motion can be understood easily within the semiclassical approach (except for damping): 
Recall that the impurity spin precession with frequency $\omega_{\rm L} \approx B_{\rm fin}$ is mainly caused by  the torque due to the magnetic field and note that the second term on the right-hand side of Eq.\ (\ref{eq:eoms}) is small if $\ff s_{i_{0}}(t)$ and $\ff S(t)$ are nearly collinear. 
In fact, in the instantaneous ground state at time $t$, the conduction-electron local moment $\ff s_{i_{0}}(t)$ would be perfectly aligned antiparallel to $\ff S(t)$ due to the antiferromagnetic exchange coupling $J$ such that $\ff s_{i_{0}}(t)$ exhibits a precessional motion with the same frequency $\omega_{\rm L} \approx B_{\rm fin}$.
Fig.\ \ref{fig:angle} demonstrates that the stronger the effective field $JS$, the smaller is the deviation of the angle $\gamma(t)$ between $\ff S(t)$ and $\ff s_{i_{0}}(t)$ from $\gamma=\pi$. 
Generally, however, $\gamma(t)<\pi$ (for all $t$) since, due to the damping, it takes a finite time for $\ff s_{i_{0}}(t)$ to react to the new position of $\ff S(t)$ (see the inset of Fig.\ \ref{fig:angle}).
Note that for very large $S$ only the time average $\overline{\gamma(t)}$ is smaller than $\pi$ (for instance, see $S \ge 20$ in Fig.\ \ref{fig:angle}).
This retardation effect results in a finite (average) torque $J \ff S(t) \times \ff s_{i_{0}}(t)$ acting on $\ff s_{i_{0}}(t)$, as can be seen from its equation of motion:
\be
\frac{d}{dt} \ff s_{i_{0}}(t)
=
J \ff S(t) \times \ff s_{i_{0}}(t)
+
T\,
\mbox{Im}
\sum_{\sigma\sigma'}
\langle c^{\dagger}_{i_{0}\sigma} \ff \tau_{\sigma\sigma'} c_{i_{0}+1\sigma'}
\rangle_{t} \: . 
\label{eq:eomss}
\ee
The second term on the right-hand side is important for energy and spin dissipation into the bulk of the system and causes the usual damping of the precession of $\ff s_{i_{0}}(t)$ (and of $\ff S(t)$) around $\ff B$. 
The first term, however, leads to nutational motion. 

This is most easily understood if there is a separation of time scales, i.e., if the nutation frequency $\omega_{\rm N}$ is large compared to the Larmor frequency $\omega_{\rm L} \approx B_{\rm fin}$.
In this limit, Eq.\ (\ref{eq:eomss}) implies that $\ff s_{i_{0}}(t)$ precesses with frequency $\omega_{N} \approx J\scl$ approximately around the momentary direction of $\ff S(t)$ (which itself slowly precesses around the field direction).
Actually, however, due to the retardation, $\ff s_{i_{0}}$ precesses around an axis which is slightly tilted as compared to 
the momentary direction of $\ff S(t)$.
This is nicely demonstrated by the oscillations of $\gamma(t)$ with time-average $\overline{\gamma(t)}<\pi$ as displayed in Fig.\ \ref{fig:angle}.
Furthermore, the equations of motion, Eq.\ (\ref{eq:eoms}) and Eq.\ (\ref{eq:eomss}), with the second term disregarded, imply that $S_{z}(t) + s_{i_{0}z}(t) = \,$const and, therefore, the impurity spin shows the same nutational motion, but with opposite amplitude.

In the middle panel of Fig.\ \ref{fig:dyn} we in fact observe a fast oscillation of $\ff s_{i_{0}}(t)$ with a frequency almost perfectly given by $J\scl$ (with $J=1$ and $S=5$). 
Note that the nutation of $\ff S(t)$ is hardly visible due to the rescaling with $S_{\rm max.}$.
The third panel for $S=50$ nicely demonstrates the nutational motion of both, $\ff s_{i_{0}}(t)$ and $\ff S(t)$, with opposite amplitudes and common frequency $\omega_{\rm N} \gg \omega_{\rm L}$.

\begin{figure}[t]
\centering
\includegraphics[width=0.9\columnwidth]{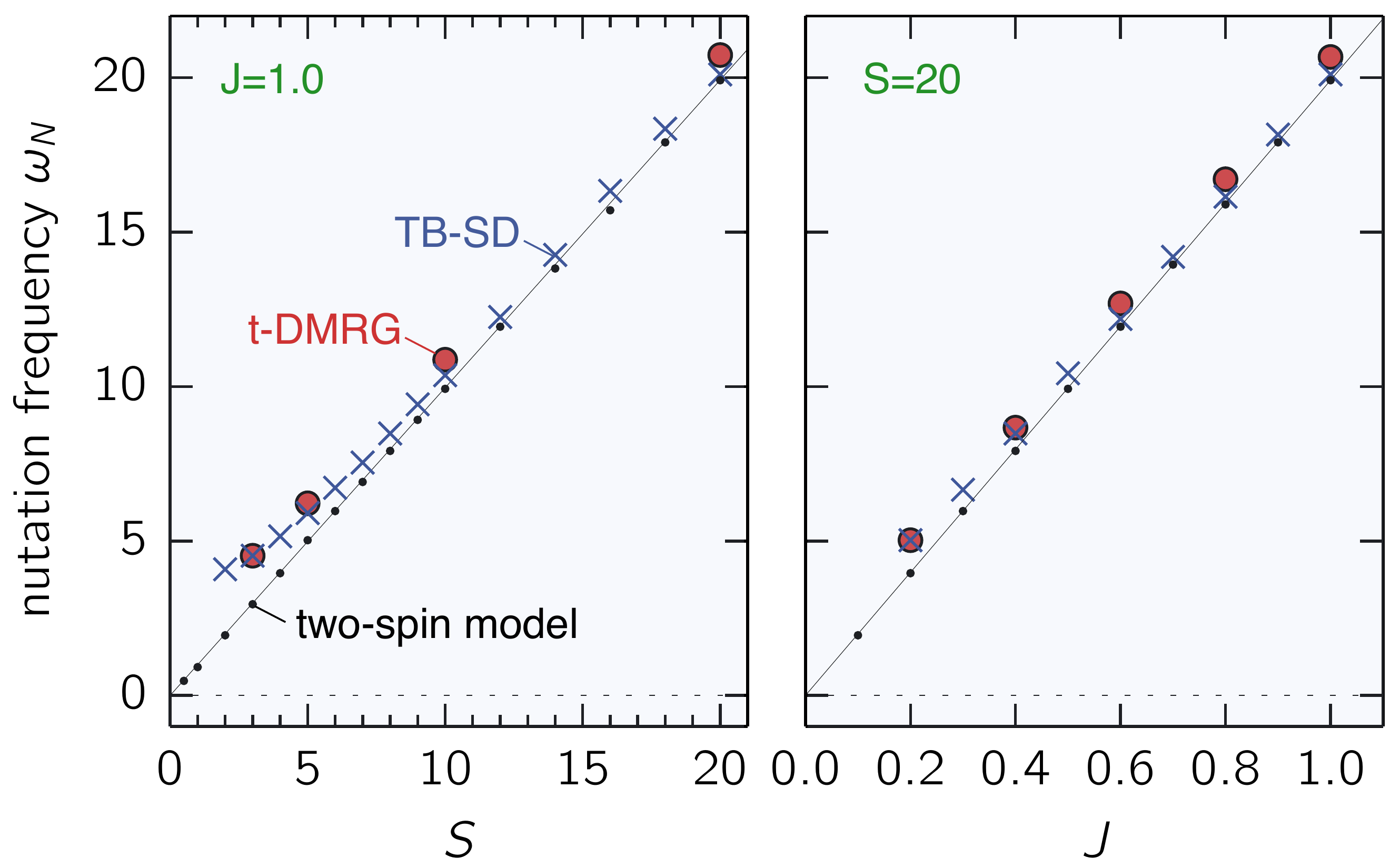}
\caption{
Nutation frequency $\omega_{\rm N}$ as a function of $S$ for $J=1$ (left) and as a function of $J$ for $S=20$ (right).
Dynamics initiated by a switch of the field from $\hat{x}$ to $\hat{z}$ direction with $B_{\rm fin} = 0.1$. 
Results for different $\scl$ or $S$, respectively, as obtained by TB-SD (crosses) and t-DMRG (circles) in comparison with the classical two-spin model (filled dots). 
} 
\label{fig:js}
\end{figure}

Fig.\ \ref{fig:js} displays the results of systematic TB-SD calculations which demonstrate the linear dependence of $\omega_{N}$ on $J$ and $S$ for large $JS$. 
These calculations have been performed for a much weaker field $B_{\rm fin} = 0.1$ resulting in a much slower precession of $\ff S(t)$ around $\ff B$. 
Note the nearly perfect agreement between classical- and quantum-spin calculations also for smaller $JS$ where there is a significant deviation from a linear behavior. 

The mechanism described above also explains that the amplitudes of the nutational oscillations vanish in the limit $S\to \infty$: 
An increasing internal Weiss field $J S$ more and more aligns $\ff s_{i_{0}}(t)$ to $\ff S(t)$, i.e., $\gamma(t) \to \pi$. 
Consequently, torque $J \ff S(t) \times \ff s_{i_{0}}(t)$ acting on $\ff s_{i_{0}}(t)$ vanishes in the large-$S$ limit.

\parag{Two-spin model.}
Fig.\ \ref{fig:js} additionally presents the results for $\omega_{\rm N}$ as obtained by a semiclassical two-spin model: 
\be
  H_{\rm 2-spin} = J \ff s \ff S - \ff B \ff S \: . 
\label{eq:toy}
\ee
This model disregards the coupling of the site $i_{0}$ to the bulk of the conduction-electron system and thus cannot describe the damping of the precessional motion.
Due to the absence of damping, the time-averaged angle is $\overline{\gamma(t)} = \pi$.

From the numerical solution of Eq.\ (\ref{eq:toy}) we also learn that it does not predict any damping of the nutational motion. 
The nutational oscillations themselves, however, are qualitatively captured by $H_{\rm 2-spin}$ and, in fact, the whole line of reasoning explaining the inertia effect also applies to this model.
The nutation frequencies as computed from $H_{\rm 2-spin}$ fit the TB-SD and t-DMRG results rather well for strong effective fields $B_{\rm eff} \equiv JS \gg T=1$; stronger deviations are found for $JS \to 2$ (see Fig.\ \ref{fig:js}). 
For $JS < 2$, there are clear nutational oscillations in the spin dynamics of the full model (\ref{eq:ham}), as is seen in the top panel of Fig.\ \ref{fig:dyn}, but $\omega_{\rm N}$ cannot be defined accurately.

\begin{figure}[t]
\centering
\includegraphics[width=0.9\columnwidth]{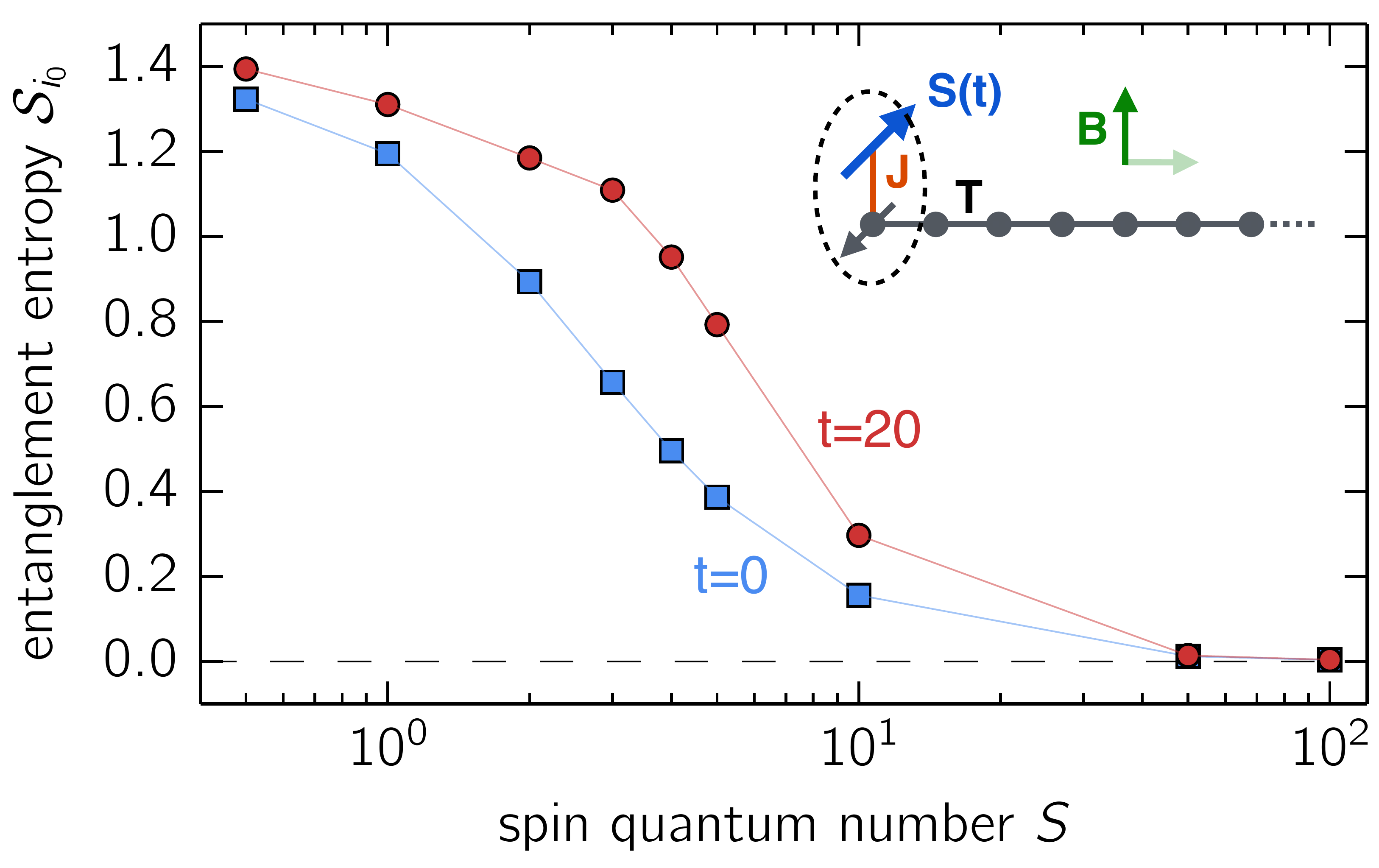}
\caption{
Entanglement entropy of the two-spin subsystem (impurity spin and site $i_{0}=1$, see dashed ellipse inset) in the environment ($i=2,...,L$) as a function of $S$ for $J=1$ and at different times $t=0$ and $t=20$.
t-DMRG results for $B_{\rm fin}=2$ and $L=50$.
} 
\label{fig:entropy}
\end{figure}

\parag{Bound states.}
$B_{\rm eff, cr} = 2$ is actually the critical value of the local effective field $B_{\rm eff} \equiv JS$ which couples to the local conduction-electron spin at $i_{0}$.
For $B_{\rm eff} > B_{\rm eff, cr}$ there are two one-particle eigenenergies of the Hamiltonian (\ref{eq:ham}) corresponding to bound states which symmetrically split off the continuum at the lower and at the upper band edge, respectively. 
Note that $B_{\rm eff, cr}$ vanishes for a site $i_{0}$ in the bulk of an infinite chain as is well known for one-dimensional systems.
Contrary, at the edge ($i_{0}=1$) there is a finite critical field, as is reminiscent of the physics in higher dimensions.

The sudden switch of the field excites the system locally at $i_{0}$. 
Consequently, if $JS > B_{\rm eff, cr}$, the subsequent dynamics is predominantly local since the excitation is mainly carried by a state whose amplitude is exponentially suppressed with increasing distance from $i_{0}$.
The dynamics should be understood in this case as a weak perturbation of the dynamics of the two-spin model Eq.\ (\ref{eq:toy}).

That this also applies to the quantum-spin case is demonstrated with Fig.\ \ref{fig:entropy} which shows the entanglement entropy $\ca S_{i_{0}}$ of the subsystem consisting of the quantum impurity spin and the conduction-electron site $i_{0}$.
In the ground state at $t=0$, the entropy decreases with increasing effective field $JS$. 
For $JS=50$ it nearly vanishes which implies that ground-state expectation values of local observables at $i_{0}$ are almost perfectly described with the (quantum version of the) two-spin model Eq.\ (\ref{eq:toy}).
With increasing time $t$, the entropy generally increases, while for strong effective fields $JS$ is stays close to zero, i.e., the two-spin model also well captures the dynamics of local observables in this case.

\parag{Damping of quantum nutation.} 
To explain the efficient damping of the nutational motion on a very short time scale $\tau_{\rm N}$ in the quantum-spin case, we first consider the quantum variant of the two-spin model Eq.\ (\ref{eq:toy}), i.e., both, $\ff S$ and $\ff s$, are considered as quantum spins with spin quantum numbers $S$ and $1/2$, respectively.  
The time-dependent expectation value $S_{z}(t)$ after the sudden switch of the field is readily computed and shows oscillations with frequency $\omega_{\rm N}$. 
Already in the two-spin model those are damped on a time scale $\tau_{\rm N}$ which agrees with that seen in the results of the full model in Fig.\ \ref{fig:dyn} for $S \ge 5$.
Writing $S_{z}(t) = \langle S_{z} \rangle_{t} = \sum_{m,n} c_{m,n} \exp(i(E_{m}-E_{n}))t$ with energy eigenstates $m$ and $n$ of $H_{\rm 2-spin}$ and coefficients $c_{m,n}$ depending on the preparation of the initial state, it becomes obvious that this damping results from the dephasing of oscillations with the excitation energies $E_{m} - E_{n}$ of the system. 

Due to the small Hilbert-space dimension of the two-spin model, however, there are strong revivals of the oscillations occurring at finite revival times. 
In fact, for $S=5$, the first revival of nutational oscillations of $s_{i_{0}z}(t)$ can be seen in the t-DMRG result around $t=20$ (Fig.\ \ref{fig:dyn}, middle panel, lower part).
With increasing $S$ and thus with increasing Hilbert space, however, the revival times quickly exceed the time scale accessible to t-DMRG in the full model. 
Furthermore, as the example for $S=5$ in Fig.\ \ref{fig:dyn} shows, the revivals themselves are strongly damped in the full theory, opposed to the nearly perfect revivals in the two-spin-model dynamics. 
As this (secondary) damping of nutation is caused by the residual effective coupling of the two-spin model to the bulk of the system, it becomes less and less efficient with increasing $S$, while at the same time the revival time strongly increases and the amplitude of the oscillations decreases.

\parag{Conclusions.} 
Inertia effects in spin dynamics have been discussed intensively in the recent years, mainly in the context of applications for magnetic devices \cite{But06,WC12,CRW11,OLW12,BH12,FSI11,BNF12,KT15,TKS08,FI11,EFC+14,KIP+09,KKR10}.
The most fundamental system which covers the essentials of spin dynamics, however, namely a single spin coupled to a Fermi sea has not yet been addressed in this respect. 
Applying exact quantum and semiclassical numerical techniques to the Kondo impurity model, we could demonstrate that the real-time dynamics, initiated by switching the direction of a magnetic field coupled to the spin, not only exhibits spin precession and spin relaxation but also nutational motion known from a gyroscope.
The effect not only shows up in the impurity-spin dynamics but also in the dynamics of the conduction-electron local magnetic moments. 
It is very robust and found in a large regime of coupling constants using tight-binding spin dynamics and treating the spin as a classical observable. 
We find that nutation amplitudes are small as compared to amplitudes in precessional motion. 
The frequency is, in the strong-coupling limit, linear in $J$ and $\scl$.

Our study has demonstrated that nutational motion is not restricted to classical-spin systems but is robust against quantum fluctuations. 
Despite the fundamental differences between semiclassical and quantum dynamics, quantum-spin nutation is found to be very similar to the classical-spin case in many respects. 
There is a qualitative, and with increasing spin-quantum numbers also quantitative agreement between quantum and semiclassical dynamics. 
Kondo screening of the impurity spin represents an important exception which, however, in the present study plays a minor role only as Kondo-singlet formation is inhibited by the external field.

The main effect of the quantum nature of the spin is a very efficient damping of the nutational motion on a very short (femtosecond) time scale which is basically independent of the relaxation time scale for the precessional motion.
In the strong-coupling ($JS \to \infty$) limit, the spin dynamics is essentially local and captured by an emergent two-spin model which has served to understand the physical origin of the damping of quantum nutation, namely dephasing of local spin excitations with revivals suppressed by the coupling to the bulk of the system. 

An important implication of our study is that direct observation of nutational motion, e.g., of magnetic nanoparticles with a (quantum)  macrospin $S$ coupled to the conduction-electron band of a nonmagnetic metallic surface, requires a sub-picosecond time resolution.
On the other hand, inertia-driven spin switching in antiferromagnets \cite{KIP+09,KKR10} has already been demonstrated successfully.

\acknowledgements

We would like to thank Christopher Stahl for instructive discussions.
Support of this work by the Deutsche Forschungsgemeinschaft within the SFB 668 (project B3), within the SFB 925 (project B5), and by the excellence cluster ``The Hamburg Centre for Ultrafast Imaging - Structure, Dynamics and Control of Matter at the Atomic Scale'' is gratefully acknowledged.
Numerical calculations were performed on the PHYSnet HPC cluster at the University of Hamburg.

\bibliographystyle{eplbib}

\begin{thebibliography}{10}
\expandafter\ifx\csname url\endcsname\relax\def\url#1{\texttt{#1}}\fi

\bibitem{Kon64}
\Name{Kondo J.} \REVIEW{Prog. Theor. Phys.}{32}{1964}{37}.

\bibitem{Hew93}
\Name{Hewson A.~C.} \Book{The Kondo Problem to Heavy Fermions} (Cambridge
  University Press, Cambridge) 1993.

\bibitem{MHK13}
\Name{Medvedyeva M., Hoffmann A. \and Kehrein S.} \REVIEW{Phys. Rev.
  B}{88}{2013}{094306}.

\bibitem{NGA+15}
\Name{Nuss M., Ganahl M., Arrigoni E., von~der Linden W. \and Evertz H.~G.}
  \REVIEW{Phys. Rev. B}{91}{2015}{085127}.

\bibitem{LL35}
\Name{Landau L.~D. \and Lifshitz E.~M.} \REVIEW{Physik. Zeits.
  Sowjetunion}{8}{1935}{153}.

\bibitem{AS06}
\Name{Anders F.~B. \and Schiller A.} \REVIEW{Phys. Rev. B}{74}{2006}{245113}.

\bibitem{SP15}
\Name{Sayad M. \and Potthoff M.} \REVIEW{New J. Phys.}{17}{2015}{113058}.

\bibitem{Gil55}
\Name{Gilbert T.} \REVIEW{Phys. Rev.}{100}{1955}{1243}.

\bibitem{Gil04}
\Name{Gilbert T.} \REVIEW{Magnetics, IEEE Transactions on}{40}{2004}{3443}.

\bibitem{Elz12}
\Name{Elze H.~T.} \REVIEW{Phys. Rev. A}{85}{2012}{052109}.

\bibitem{Sal12}
\Name{Salcedo L.~L.} \REVIEW{Phys. Rev. A}{85}{2012}{022127}.

\bibitem{SRP16}
\Name{Sayad M., Rausch R. \and Potthoff M.} \REVIEW{Phys. Rev.
  Lett.}{117}{2016}{127201}.

\bibitem{Sch11}
\Name{Schollw\"ock U.} \REVIEW{Ann. Phys. (N.Y.)}{326}{2011}{96}.

\bibitem{HCO+11}
\Name{Haegeman J., Cirac J.~I., Osborne T.~J., Pi{\v{z}}orn I., Verschelde H.
  \and Verstraete F.} \REVIEW{Phys. Rev. Lett.}{107}{2011}{070601}.

\bibitem{But06}
\Name{Butikov E.} \REVIEW{European Journal of Physics}{27}{2006}{1071}.

\bibitem{WC12}
\Name{Wegrowe J.-E. \and Ciornei M.-C.} \REVIEW{Am. J. Phys.}{80}{2012}{607}.

\bibitem{CRW11}
\Name{Ciornei M.-C., Rub\'{i} J.~M. \and Wegrowe J.-E.} \REVIEW{Phys. Rev.
  B}{83}{2011}{020410}.

\bibitem{OLW12}
\Name{Olive E., Lansac Y. \and Wegrowe J.-E.} \REVIEW{Appl. Phys.
  Lett}{100}{2012}{192407}.

\bibitem{BH12}
\Name{B\"ottcher D. \and Henk J.} \REVIEW{Phys. Rev. B}{86}{2012}{020404(R)}.

\bibitem{FSI11}
\Name{F\"ahnle M., Steiauf D. \and Illg C.} \REVIEW{Phys. Rev.
  B}{84}{2011}{172403}.

\bibitem{BNF12}
\Name{Bhattacharjee S., Nordstr\"om L. \and Fransson J.} \REVIEW{Phys. Rev.
  Lett.}{108}{2012}{057204}.

\bibitem{KT15}
\Name{Kikuchi T. \and Tatara G.} \REVIEW{Phys. Rev. B}{92}{2015}{184410}.

\bibitem{TKS08}
\Name{Tatara G., Kohno H. \and Shibata J.} \REVIEW{Physics
  Reports}{468}{2008}{213}.

\bibitem{FI11}
\Name{F\"ahnle M. \and Illg C.} \REVIEW{J. Phys.: Condens.
  Matter}{23}{2011}{493201}.

\bibitem{EFC+14}
\Name{Evans R. F.~L., Fan W.~J., Chureemart P., Ostler T.~A., Ellis M. O.~A.
  \and Chantrell R.~W.} \REVIEW{J. Phys.: Condens. Matter}{26}{2014}{103202}.

\bibitem{KIP+09}
\Name{Kimel A.~V., Ivanov B.~A., Pisarev R.~V., Usachev P.~A., Kirilyuk A. \and
  Rasing T.} \REVIEW{Nat. Physics}{5}{2009}{727}.

\bibitem{KKR10}
\Name{Kirilyuk A., Kimel A.~V. \and Rasing T.} \REVIEW{Rev. Mod.
  Phys.}{82}{2010}{2731}.

\bibitem{GvdBH+02}
\Name{Gerrits T., van~den Berg H. A.~M., Hohlfeld J., B\"ar L. \and Rasing T.}
  \REVIEW{Nature (London)}{418}{2002}{509}.

\bibitem{TSK+04}
\Name{Tudosa I., Stamm C., Kashuba A.~B., King F., Siegmann H.~C., St\"ohr J.,
  Ju G., Lu B. \and Weller D.} \REVIEW{Nature (London)}{428}{2004}{831}.

\bibitem{Wie09}
\Name{Wiesendanger R.} \REVIEW{Rev. Mod. Phys.}{81}{2009}{1495}.

\bibitem{NF93}
\Name{Nunes G. \and Freeman M.~R.} \REVIEW{Science}{262}{1993}{1029}.

\bibitem{Mor10}
\Name{Morgenstern M.} \REVIEW{Science}{329}{2010}{1609}.

\bibitem{LEL+10}
\Name{Loth S., Etzkorn M., Lutz C.~P., Eigler D.~M. \and Heinrich A.~J.}
  \REVIEW{Science}{329}{2010}{1628}.

\bibitem{YCB+14}
\Name{Yan S., Choi D.-J., Burgess J. A.~J., Rolf-Pissarczyk S. \and Loth S.}
  \REVIEW{Nat. Nanotechnol.}{10}{2015}{40}.

\bibitem{HLO+14}
\Name{Haegeman J., Lubich C., Oseledets I., Vandereycken B. \and Verstraete F.}
  \REVIEW{}{}{2014}{}.

\bibitem{LR72}
\Name{Lieb E.~H. \and Robinson D.~W.} \REVIEW{Commun. Math.
  Phys.}{28}{1972}{251}.

\bibitem{BHV06}
\Name{Bravyi S., Hastings M.~B. \and Verstraete F.} \REVIEW{Phys. Rev.
  Lett.}{97}{2006}{050401}.

\bibitem{Ver10}
\Name{Verner J.~H.} \REVIEW{Numerical Algorithms}{53}{2010}{383}.

\bibitem{Lie73}
\Name{Lieb E.~H.} \REVIEW{Commun. Math. Phys.}{31}{1973}{327}.

\bibitem{KFG99}
\Name{Kladko K., Fulde P. \and Garanin D.~A.} \REVIEW{Europhys.
  Lett.}{46}{1999}{425}.

\bibitem{GKF00}
\Name{Garanin D.~A., Kladko K. \and Fulde P.} \REVIEW{Euro. Phys. J.
  B}{14}{2000}{293}.

\bibitem{Gar08}
\Name{Garanin D.~A.} \REVIEW{Phys. Rev. B}{78}{2008}{144413}.

\bibitem{SGP12b}
\Name{Sayad M., G\"utersloh D. \and Potthoff M.} \REVIEW{Euro. Phys. J.
  B}{85}{2012}{125}.

\bibitem{suppl}
See Supplemental Material at {\em URL}.

\bibitem{BF02}
\Name{Breuer H.~P. \and Petruccione F.} \Book{The Theory of Open Quantum
  Systems} (Oxford University Press, New York) 2002.

\bibitem{LLG}
L.~D. Landau and E.~M. Lifshitz, Physik. Zeits. Sowjetunion \textbf{8},153
  (1935); T. Gilbert, Phys. Rev. \textbf{100}, 1243 (1955); T. Gilbert,
  Magnetics, IEEE Transactions on \textbf{40}, 3443 (2004).

\bibitem{Kik56}
\Name{Kikuchi R.} \REVIEW{J. Appl. Phys.}{27}{1956}{1352}.

\end{thebibliography}


\newpage

\begin{center}

{\large \bfseries
Inertia effects in quantum and classical dynamics of a spin coupled to a Fermi sea
\\ \mbox{} \\
--- Supplemental material --- 
}

\mbox{} \\

Mohammad Sayad, Roman Rausch and Michael Potthoff \\

{\em \small I. Institut f\"ur Theoretische Physik, Universit\"at Hamburg, Jungiusstra\ss{}e 9, 20355 Hamburg, Germany}
\end{center}

\parag{Classical spin-only theory.}
It is instructive to discuss the $\scl$ dependence of the spin damping and nutation in an effective classical spin-only theory, i.e., after integrating out the electron degrees of freedom.
Following Ref.\ \cite{BNF12}, an effective equation of motion for $\ff S(t)$ is obtained for the classical-spin case by (i) lowest-order perturbation theory in $J$ and (ii) assuming that the spin dynamics is slow: 

(i) In the weak-$J$ regime, we can use the Kubo formula to find the linear response $\ff s_{i_{0}}(t) =  J \int_{0}^{t} dt' \, {\chi_{\rm loc}}(t-t') \ff S(t')$ of the local conduction-electron magnetic moment at site $i_{0}$ and time $t$ caused by the time-dependent effective field $\ff B_{\rm eff}(t') \equiv J \ff S(t')$ at time $t'$.
Here, the time-homogeneous response function $\chi_{\rm loc}(t-t')$ is the retarded local spin susceptibility of the electron system at $i_{0}$. 
This is a rank-two tensor which, for the present case, is diagonal and constant: 
$\chi_{\rm loc}(t) = - i \Theta(t) \langle 0 | [s_{i_{0}z}(t) , s_{i_{0}z}(0)] | 0 \rangle$ where $| 0 \rangle$ is the initial ground state at time $t=0$, where $\Theta$ is the Heaviside step function and where $s_{i_{0}z}(t) = \exp(iH_{\rm e}t) s_{i_{0}z}(0) \exp(-iH_{\rm e}t)$ with the tight-binding Hamiltonian $H_{\rm e}$ [first term in Eq.\ (\ref{eq:ham})].
Inserting into Eq.\ (\ref{eq:eoms}), we find
\begin{equation}
\dot{\ff S}(t)
= 
\ff S(t) \times \ff B \\
- J^{2} \ff S(t) \times \int_{0}^{t} dt' \, \chi_{\rm loc}(t-t') \ff S(t') \; .
\label{eq:id}   
\end{equation}

(ii) Assuming that the classical spin is slow on the memory time scale set by $\chi_{\rm loc}(t)$ and expanding 
$\ff S(t') = \ff S(t) + \dot{\ff S}(t) (t'-t) + \ddot{\ff S}(t') (t-t')^{2}/2 + \cdots$ under the integral, one finds \cite{BNF12} the LLG equation with an additional inertia term: 
\be
\dot{\ff S} = \ff S \times \ff B - \alpha \ff S \times \dot{\ff S} + I \ff S \times \ddot{\ff S} \: , 
\label{eq:llg+}
\ee
where, after sending the upper integral limit to infinity as usual \cite{BF02,BNF12,SP15}, 
\be
\alpha = - J^{2} \int_{0}^{\infty} d\tau \, \tau \chi_{\rm loc}(\tau)
\label{eq:ai1}
\ee
and 
\be
I = - \frac{J^2}{2} \int_{0}^{\infty} d\tau \, \tau^{2} \chi_{\rm loc}(\tau)
\label{eq:ai2}
\ee
are the Gilbert damping constant and the moment of inertia, respectively ($\alpha, I >0$).
Eq.\ (\ref{eq:llg+}) constitutes a purely classical spin-only theory which, after some extensions, can serve as a starting point for microscopic spin-dynamics calculations \cite{LLG}.
The inertia term is known to give rise to nutation (see, e.g., \cite{BNF12,KT15} for a detailed discussion).

\parag{Dependence on $\scl$.}
The equation of motion (\ref{eq:llg+}) has a simple scaling property:
One can easily verify that if $\ff S(t)$ solves the equation for parameters $\ff B, \alpha$ and $I$, then $\ff S'(t) \equiv \lambda \ff S(t)$ solves the same equation with rescaled parameters $\alpha / \lambda$ and $I/\lambda$.
We conclude that the damping parameter $\alpha$ and the inertia constant $I$ have a stronger effect on the dynamics of an elongated spin ($\scl >1$). 
Namely, with smaller effective parameters
\be
  \alpha' = \alpha / \scl \; , \qquad I' = I / \scl \: ,
\ee  
one obtains the same dynamics as for a spin of unit length.

As is obvious from the defining equations (\ref{eq:ai1}) and (\ref{eq:ai2}), the parameters $\alpha, I$ do not depend on $\scl$ but are properties of the conduction-electron system only. 
The equation of motion for a given system is therefore independent of $\scl$.

For fixed $\alpha$, the relaxation time can be calculated \cite{Kik56} and is given by $\tau_{\rm rel} \propto (1+\alpha^{2} \scl^{2}) / (\alpha \scl B) $.
In the large-$\scl$ limit, we thus have
\be
  \tau_{\rm rel} \propto \scl \: .
\label{eq:tau}
\ee

Identifying $\scl$ with the modulus of the angular momentum $L$ of a fast-spinning gyroscope, elementary theory (see, e.g., Ref.\ \cite{But06}) tells us $L = I \omega_{\rm N}$, and hence
\be
  \omega_{\rm N} \propto \scl \: .
\label{eq:nut}
\ee
This recovers the numerical result found in the $\scl \to \infty$ limit and corroborates the interpretation given in the main text.
It also appears more general and does not depend on the special form of the underlying equations of motion as long as the motion of a magnetic moment is concerned \cite{BNF12,KT15}.

\parag{Discussion.}
In fact, Eq.\ (\ref{eq:nut}), is almost perfectly verified within the framework of the full semiclassical TB-SD, see Fig.\ \ref{fig:js} for large $\scl$.
For the dependence of the relaxation time $\tau_{\rm rel}$ on $\scl$, however, we find a quadratic relation in the full TB-SD, $\tau_{\rm rel} \propto \scl^{2}$, rather than the linear trend predicted by Eq.\ (\ref{eq:tau}).
This indicates that the effective theory is of limited use for reproducing the exact results of TB-SD on the semiclassical level for large $\scl$ and is furthermore inconsistent with the quantum dynamics as well.

Both assumptions (i) and (ii) are in fact questionable for the parameter regime studied here:
One may reject the Markov-type approximation (ii) and describe the spin dynamics with the linear-response theory and Eq.\ (\ref{eq:id}).
The same scaling argument as above again tells us that an elongated ($\lambda>1$) spin $\ff S'(t) \equiv \lambda \ff S(t)$ solves the same integro-differential equation, 
\be
  \ff S'(t,\ff B, J') = \ff S(t,\ff B, J) \; , 
\ee
but with rescaled exchange coupling 
\be
  J'=J/\sqrt{\lambda} \: .
\label{eq:scale}  
\ee
This means that a weaker interaction $J' < J$ (for $\lambda>1$) leads to the same dynamics. 

Now, from the numerical evaluation of Eq.\ (\ref{eq:id}) it is well known \cite{SP15,SRP16} that, for fixed $\scl$, the damping becomes stronger and the relaxation time shorter with increasing $J$.
Hence, for fixed $J$, with the argument leading to Eq.\ (\ref{eq:scale}), the dynamics of an elongated classical spin must therefore show a stronger damping, i.e., a shorter relaxation time.

As this conflicts with our observations here (see Fig.\ \ref{fig:dyn}), we must conclude that approximation (i), i.e., lowest-order perturbation theory in $J$ or linear-response theory, is no longer valid for the parameter regime studied here.
This furthermore implies that, besides damping, also the nutational motion of a spin exchange coupled to an unpolarized Fermi sea cannot be captured by the perturbative approach (despite the fact that the linear trend Eq.\ (\ref{eq:nut}) is reproduced).
This is not too surprising in view of the explanation for the inertia effect given in the main text, namely the formation of a bound state of the impurity spin with the exchange-coupled conduction-electron spin and the weak interaction of this bound state with the bulk of the system.
Those details of the electronic structure are obviously not accounted for in a simple effective spin-only theory, such as Eq.\ (\ref{eq:id}), where the electron dynamics only enters via the $J=0$ spin susceptibility.

\end{document}